\documentclass[showpacs,preprintnumbers,amsmath,amssymb,twocolumn, prl]{revtex4}
\usepackage{amsfonts}
\usepackage{mathrsfs}
\usepackage{graphicx}% Include figure files
\usepackage{dcolumn}% Align table columns on decimal point
\usepackage{bm}% bold math

\begin{document}

\title{Unified Universal Quantum Cloning Machine and Fidelities}

\author{Yi-Nan Wang$^1$, Han-Duo Shi$^1$, Zhao-Xi Xiong$^1$, Li Jing$^1$,
Xi-Jun Ren$^2$, Liang-Zhu Mu$^1$\footnote{muliangzhu@pku.edu.cn},
and Heng Fan$^3$\footnote{hfan@iphy.ac.cn}}
\affiliation{%
$^1$School of Physics, Peking University, Beijing 100871, China\\
$^2$School of Physics and Electronics, Henan University, Kaifeng 4750011, China\\
$^3$Institute of Physics, Chinese Academy of Sciences, Beijing
100190, China
}%
\date{\today}% It is always \today, today,

\begin{abstract}
We present a unified universal quantum cloning machine, which
combines several different existing universal cloning machines
together including the asymmetric case. In this unified framework,
the identical pure states are projected equally into each copy
initially constituted by input and one half of the maximally
entangled states. We show explicitly that the output states of those
universal cloning machines are the same. One importance of this
unified cloning machine is that the cloning procession is always the
symmetric projection which reduces dramatically the difficulties for
implementation. Also it is found that this unified cloning machine
can be directly modified to the general asymmetric case. Besides the
global fidelity and the single-copy fidelity, we also present all
possible arbitrary-copy fidelities.
\end{abstract}

\maketitle

\emph{Introduction}.--- No-cloning theorem is fundamental for
quantum mechanics and quantum information science that states an
unknown quantum state can not be cloned perfectly\cite{noncloning}.
However, we can try to clone a quantum state approximately with the
optimal quality \cite{UQCM}, or instead, we can try to clone it
perfectly with the largest probability \cite{DG}. So various quantum
cloning machines have been designed for different quantum
information tasks [4-18].
%For
%identical unknown or arbitrary quantum states, we can use universal
%quantum cloning machine which creates several identical copies of
%the input, the quality of the copies is independent of the input.
%If
%we know {\it a priori} partial information of the input state, we
%can use the phase-covariant quantum cloning machine. For instance in
%quantum cryptography protocol, the well-known BB84 states can be
%optimally cloned by the phase-covariant quantum cloning machine.
%Recently much progress has been made in studying quantum cloning and
%related topics.
Experimentally, quantum cloning machines have been realized in
optics system \cite{clonescience02,naturephotonics09,fs,cv}, nuclear
magnetic resonance system \cite{Du,Chendupra07}, diamond
nitrogen-vacancy center system \cite{PanFan}, etc.

The universal quantum cloning machine is first proposed by Bu\v{z}ek
and Hillery \cite{UQCM} which can copy optimally one arbitrary qubit
equally well to two copies. Later more general cases have been
studied, see Ref.\cite{rmp} for a review. So far there exists two
universal quantum cloning machines which can clone $N$ identical
$d$-level pure quantum states to $M$ copies, $M\ge N$. One is
proposed by Werner in Ref. \cite{Werner}, and the other is proposed
by Fan \emph{et al.} in Ref. \cite{Fan}.
%The advantage of Werner
%cloning machine is that the transformation is concise and easily
%understood, i.e., the pure input state is symmetrically distributed
%to each copy. While the cloning machine given in Ref. \cite{Fan}
%provides explicitly the unitary transformation with ancillary states
%so that the quantum circuit can be designed correspondingly. And it
%is shown that this cloning machine might be realized by emission of
%light in a multilevel atomic system \cite{Fanphysical}.
%Both are optimal, however, the
%explicit equivalence of these two cloning machines has not yet been
%presented. In this Letter, we will first show that the density
%operators from the two cloning machines are the same.
Both have advantages from different points of view.  Also we know
some limited cases of asymmetric cloning \cite{Cerf}. It seems that
all those cloning machines are quite different and no simple
connection exists. In this Letter, we will present a simple and
unified cloning transformation which combines all those cloning
machines together. In the framework of the unified quantum cloning,
the identical pure states are projected equally into each copy where
the output is initially constituted by input and one half of the
maximally entangled states, the left half of the maximally entangled
states act as the ancillary states. We will show explicitly that the
density operators from those cloning machines are the same. Since
the symmetric operator is constituted by SWAP gates, the quantum
circuit corresponding to this unified cloning machine can thus be
designed accordingly. Also importantly, this unified universal
cloning machine can be easily modified to the general asymmetric
case. As we know, only a few limited results of asymmetric cloning
are known while the general case is still absent.

The optimality of the cloning machine is judged generally by whether
the obtained fidelity of the cloning output state achieves its
optimal bound. So far the optimal global fidelity and single-copy
fidelity have already been obtained \cite{Werner,KW}. It is,
however, necessary to find optimal general-copy fidelities since it
is possible that different optimal fidelities give different
criteria, as happened in phase-covariant cloning machine
\cite{Fanjpa}. With the unified and optimal universal cloning
machine, we will present all possible arbitrary-copy fidelities so
that the optimality of the cloning machine can be quantified from
different aspects.

\emph{Equivalence of two universal quantum cloning machines}.---For a pure state, $|\varphi \rangle =\sum _jx_j|j\rangle $,
in $d$-dimensional Hilbert space $\mathcal{H}$ with density operator $\sigma \equiv |\varphi \rangle \langle \varphi|$,
$d$=dim$\mathcal{H}$, $\sum _j|x_j|^2=1$, we refer it to a qudit.
The $N\rightarrow M$ Werner cloning machine is presented as \cite{Werner},
\begin{eqnarray}
\rho ^{out}=\frac {d[N]}{d[M]}s_M\left( \sigma ^N\otimes {\mathbb{I}}^{\otimes (M-N)}\right)s_M,
\label{Werner}
\end{eqnarray}
where $d[N]=C_{d+N-1}^N=\frac {(d+N-1)!}{N!(d-1)!}$, it is the dimension of the symmetric subspace of $N$-fold Hilbert space,
$d[N]={\rm dim}\mathcal{H}^{\otimes N}_+$, ${\mathbb{I}}$ is the identity on $\mathcal{H}$,
$s_M$ is the symmetric projector which maps states in $\mathcal{H}^{\otimes M}$ onto its symmetric
subspace $\mathcal{H}^{\otimes M}_+$. Explicitly, $s_M=\sum _{\vec {m}}^M|\vec {m}\rangle \langle \vec {m}|$, where
state $|\vec {m}\rangle \equiv |m_1,m_2,...,m_d\rangle $ is a completely symmetric state with $m_j$ states in $|j\rangle $,
and the summation is assumed to run all possible values with constraint, $\sum _jm_j=M$.

In Werner cloning machine, the information of the input is equally
projected to each copy. The initial $M-N$ identities are also understandable since
the assumption of the universal cloning machine is that the input state is arbitrary
thus a completely mixed state, $\mathbb{I}/d$, should be a suitable candidate
before the cloning procession is applied.

Following Bu\v{z}ek-Hillery $1\rightarrow 2$ universal cloning
machine for qubit \cite{UQCM} and qudit \cite{d_dim}, the
$N\rightarrow M$ for qubit by Gisin and Massar \cite{N_UQCM}, Fan
\emph{et al.} proposed the following transformation for qudit case
\cite{Fan}, see Fig.1,
\begin{eqnarray}
U|\vec {n}\rangle \otimes R=\eta \sum _{\vec {k}}^{M-N}\sqrt {{\prod _j}\frac {(n_j+k_j)!}{n_j!k_j!}}
|\vec {n}+\vec {k}\rangle \otimes R_{\vec {k}},
\label{Fanclone}
\end{eqnarray}
where $|\vec {n}\rangle $ is the input state, $U$ denotes unitary transformation,
$R$ in l.h.s. denotes blank state and the initial ancillary state, $R_{\vec {k}}$ is the
ancillary state which can be realized also by symmetric state $|\vec {k}\rangle $, the
whole normalization factor takes the form, $\eta =\sqrt {(M-N)!(N+d-1)!/(M+d-1)!}$. Here,
the summation is also assumed to run all possible values under the constraint $\sum k_j=M-N$.
For $N$ identical pure states $|\varphi \rangle $, the input takes the form,
\begin{eqnarray}
|\varphi \rangle ^{\otimes N}=\sqrt {N!}\sum _{\vec {n}}^N\prod _j\frac {x_j^{n_j}}
{\sqrt {n_j!}}|\vec {n}\rangle ,
\label{expansion}
\end{eqnarray}
we can apply Eq. (\ref{Fanclone}), trace out the ancillary states
and thus obtain the output density operator. Similarly as for qubit
case in theory \cite{Emission} and in experiment
\cite{clonescience02}, this cloning machine might naturally be
realized by light emission from multilevel atomic system
\cite{Fanphysical}.

Typically, two different figures of merit are applied for the universal cloning machines.
One is the global fidelity between the whole output density operator $\rho ^{out}$ and the
ideal output $|\varphi \rangle ^{\otimes M}$ with perfect $M$ copies,
$F_M=^{\otimes M}\langle \varphi|\rho ^{out}|\varphi \rangle ^{\otimes M}$. The other is the single-copy
fidelity defined between an individual output density operator and a single input pure state,
$F_1=\langle \varphi|\rho ^{out}_1|\varphi \rangle $, where each individual output density
operator $\rho ^{out}_1$ is the same for all $M$ copies. It is shown that those two fidelities
for Werner cloning machine and for cloning machine in (\ref{Fanclone}) are the same. Yet the
similarity of the output density operators is necessary.
Next, explicitly we shall show the output state of $M$ copies from the two cloning machines
are the same.

First, we can find that the symmetric state $|\vec {m}\rangle $ of $M$ qudits
can be divided into two parts with N qudits and $M-N$ qudits, respectively,
\begin{eqnarray}
|\vec {m}\rangle =\frac {1}{\sqrt {C_M^N}}\sum _{\vec {k}}^{M-N}\prod _j\sqrt {\frac {m_j!}{(m_j-k_j)!k_j!}}|\vec {m}-\vec {k}\rangle
|\vec {k}\rangle .
\label{splitting}
\end{eqnarray}
The symmetric projector $s_M$ can thus be reformulated by this
splitting, then with the help of the expansion (\ref{expansion}),
substituting $\sigma ^{\otimes N}$  into (\ref{Werner}), by some
calculations, the output density operator of Werner cloning machine
takes the form,
\begin{eqnarray}
\rho ^{out}
%&=&\frac {d[N]N!}{d[M]}\sum _{\vec {m},\vec {m}',\vec {n},\vec {n}',\vec {k},\vec {k}'}
%|\vec {m}\rangle \langle \vec {m}|\times
%\nonumber \\
%&&\times \left(
%\prod _{j}\frac {x_j^{n_j}x_j^{*n_j'}\sqrt {m_j!m_j'!}}{\sqrt {(m_j-k_j)!(m_j'-k_j')!k_j!k'_j!n_j!n_j'!}}\right) \times
%\nonumber \\
%&&\times
%\left[
%\langle \vec {m}-\vec {k}|\langle \vec {k}\right] \left[ (|\vec {n}\rangle \langle \vec {n}')\otimes {\mathbb{I}}^{(M-N)}\right]
%\left[|\vec {m}'-\vec {k}'\rangle |\vec {k}'\rangle \right]
%\nonumber \\
&=&N!\eta ^2\sum _{\vec {m},\vec {m}'}^M|\vec {m}\rangle \langle
\vec {m}'|
\nonumber \\
&&\times \left(
\sum _{\vec {k}}^{M-N}\prod _j\frac {x_j^{m_j-k_j}x^{*(m_j'-k_j)}\sqrt {m_j!m_j'!}}
{(m_j-k_j)!(m'_j-k_j)!k_j!}\right),
\end{eqnarray}
where we have already used $\langle \vec {l}|\vec {l}'\rangle =\delta _{\vec {l}\vec {l}'}$.

For the second cloning machine (\ref{Fanclone}), substituting the result of (\ref{expansion}) into (\ref{Fanclone})
and tracing out the ancillary states, simply the output state is written as
\begin{eqnarray}
{\rho '}^{out}&=&N!\eta ^2\sum _{\vec {n},\vec {n}'}^N\sum _{\vec {k}}^{M-N}
|\vec {n}+\vec {k}\rangle \langle \vec {n}'+\vec {k}|
\nonumber \\
&&\times \left(
\prod _j\frac {x_j^{n_j}x_j^{*(n'_j)}\sqrt {(n_j+k_j)!(n_j'+k_j)!}}{n_j!n_j'!k_j!}\right).
\end{eqnarray}
Considering we have the constraint $\sum k_j=M-N$, apparently, the output states from
two universal quantum cloning machines are the same, ${\rho }^{out}={\rho '}^{out}$.

\begin{figure}
\includegraphics[height=3cm,width=6.5cm]{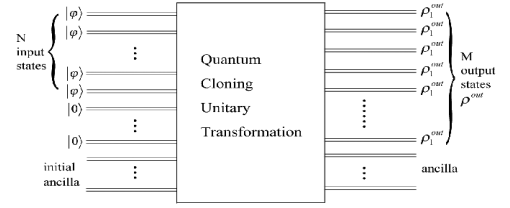}
\caption{Generally, the input of the cloning machines are identical
pure input states, blank states and initial ancilla, the cloning is
realized by unitary transformation.}
\end{figure}

\emph{Unified universal quantum cloning machine}.---Stimulated by the fact that the two existing
universal cloning machines are the same, we may try further to explore the possibility to unify the two
cloning machines together. For $N$ identical pure states, we propose
the $N\rightarrow M$ universal cloning machine as the following,
\begin{eqnarray}
|\varphi \rangle ^{out}=\lambda \left( s_{M}\otimes {\mathbb{I}}^{\otimes (M-N)}\right) |\varphi \rangle ^{\otimes N}|\Phi ^+\rangle ^{\otimes (M-N)},
\label{unified}
\end{eqnarray}
where $|\Phi ^+\rangle \equiv \frac {1}{\sqrt {d}}\sum _j|jj\rangle
$ is a maximally entangled state in $\mathcal{H}\otimes
\mathcal{H}$, $\lambda $ is the normalization factor. Here operator
$s_M$ acts on the $N$ identical pure input states and half of the
$M-N$ maximally entangled states, the left half of the maximally
entangled states are the ancillary states, see Fig.2.

Next we will show that this cloning transformation is the same as
(\ref{Fanclone}). We consider a property of the symmetric projector,
$s_M=s_M\left( {\mathbb{I}}^{\otimes N}\otimes s_{M-N}\right) $,
this is due to Eq.(\ref{splitting}) and also $s_{M-N}$ is the
identity operator on $\mathcal{H}^{\otimes (M-N)}_+$. By symmetric
projection, $M-N$ maximally entangled states can be mapped as a
maximally entangled state in symmetric subspace
$\mathcal{H}^{\otimes (M-N)}_+\otimes \mathcal{H}^{\otimes
(M-N)}_+$,
\begin{eqnarray}
\left( s_{M-N}\otimes {\mathbb{I}}^{\otimes (M-N)}\right) |\Phi ^+\rangle ^{\otimes (M-N)}=
\sum _{\vec {k}}^{M-N}|\vec {k}\rangle |\vec {k}\rangle ,
\label{schmidt}
\end{eqnarray}
where an unimportant whole factor is omitted, and some other unimportant whole factors
will also be omitted later without specification.
Note that the entanglement cutting is unchanged here.
Since quantum mechanics is linear,
so we just consider the input state be a symmetric state for (\ref{unified}), with the help
of the result in (\ref{schmidt}), we can find the unified cloning machine (\ref{unified}) can
be rewritten as the form
\begin{eqnarray}
&&\left( s_{M}\otimes {\mathbb{I}}^{\otimes (M-N)}\right) |\vec {n}\rangle |\Phi ^+\rangle ^{\otimes (M-N)}
\nonumber \\
&&=\left( s_{M}\otimes {\mathbb{I}}^{\otimes (M-N)}\right)
|\vec {n}\rangle \sum _{\vec {k}}^{M-N}|\vec {k}\rangle |\vec {k}\rangle
\nonumber \\
&&=\sum _{\vec {k}}^{M-N}\sqrt {\prod _j\frac {(n_j+k_j)!}{n_j!k_j!}}|\vec {n}+\vec {k}\rangle
|\vec {k}\rangle ,
\end{eqnarray}
where the splitting relation (\ref{splitting}) is used in the
last equation.
Thus considering that the last $M-N$ qudits are ancillary states,
the universal cloning transformation (\ref{Fanclone}) is re-obtained by the unified cloning transformation
(\ref{unified}). Also (\ref{unified}) can be considered to be an equivalent form of Werner cloning machine since
by taking trace over the last $M-N$ qudits where each maximally entangled state will provide
an identity on $\mathcal{H}$, we will re-obtain Werner cloning machine (\ref{Werner}).

We remark that the unified cloning machine (\ref{unified}) can be
easily understood, which is a property inherited from Werner cloning
machine, and it has also the explicit transformations as those in,
such as Refs.\cite{UQCM,N_UQCM,d_dim,Fan}.

\begin{figure}
\includegraphics[height=3cm,width=6cm]{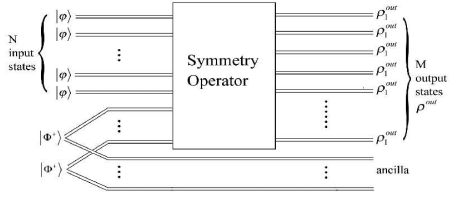}
\vskip 0.3truecm \caption{The unified cloning machine is constituted
by identical pure input states and the prepared maximally entangled
states, the cloning is always realized by symmetric projection.}
\end{figure}

\emph{General fidelities}.---The merit of the cloning machine is
generally quantified by fidelity between input and output, the
global fidelity and single-copy fidelity are presently known. As we
mentioned, it is also necessary to have all possible arbitrary-copy
fidelities to offer a full description of the merit of universal
cloning machine. With output density operator $\rho ^{out}$
available, we can find the reduced density operators of $L$ qudits,
$\rho ^{out}_L$, where $1\le L\le M$. The general fidelities are
defined as $F_L\equiv ^{\otimes L}\langle \varphi |\rho
^{out}_L|\varphi \rangle ^{\otimes L}.$ We remark that all reduced
density operators of $L$ qudits at different positions are the same
which is ensured by the fact that the output is in symmetric
subspace ${\mathcal {H}}^{\otimes M}_+$. By straightforward but
tedious calculations, we find,
\begin{eqnarray}
&&F_L=\frac {(d+N-1)!(M-N)!(M-L)!}{(d+M-1)!M!N!}\times
\nonumber \\
&&\sum _{m_1}\frac
{(M-m_1+d-2)!(m_1!)^2}{(m_1-L)!(m_1-N)!(d-2)!(M-m_1)!},
\label{general}
\end{eqnarray}
where $m_1$ is one entry of the vector $\vec {m}$, here we have considered
the property that the output state $\rho ^{out}$ is covariant for the cloning transformation,
i.e., $\rho ^{out}$ is changed as $u^{\otimes M}\rho ^{out}u^{\dagger \otimes M}$ when
$|\varphi \rangle  $ is changed as $u|\varphi \rangle $ \cite{Werner}.

For $L=1,L=M$, we recover the known results \cite{Werner,KW}.
$F_M={d[N]}/{d[M]}$, and
$F_1=({N(d+M)+M-N})/{(d+N)M}.$

For special case $N=1$, the general fidelity (\ref{general}) can be
simplified as,
\begin{eqnarray}
F_L(N=1)=\frac {L!d![L(d+M)+M-L] }{(d+L)!M}.
\end{eqnarray}
Here we have used the identity:
\begin{eqnarray}
&&\frac {(M-N)!(N+d-1)!}{(M+d-1)!N!}\times
\nonumber \\
&&
\sum _{m=0}^{M-N}\frac {((N+m)!)^2(M-N-m+d-2)!}{Mm(M-N-m)!(d-2)!}
\nonumber \\
&&
=\frac {N(d+M)+M-N}{(d+N)M}\nonumber
\end{eqnarray}
The proof of this identity is mainly based on a permutation and
combination equation \cite{combination}.

\emph{Extension of the unified cloning machine to asymmetric case and examples}.---As an example,
let us consider the 1 to 2 cloning machine for qudit and qubit. The symmetric
projector takes the form $s_2=\sum |jj\rangle \langle jj|+\frac
{1}{2}\sum _{j\not =l}(|jl\rangle +|lj\rangle )(\langle jl|+\langle
lj|)$. Up to a whole factor, the unified cloning machine can be written as,
\begin{eqnarray}
&&(s_2\otimes {\mathbb{I}} )|l\rangle _1|\Phi ^+\rangle _{2a}=
|ll\rangle _{12}|l\rangle _a
\nonumber \\
&&~~~+\frac {1}{2}\sum _{j\not =i}(|lj\rangle _{12}+|jl\rangle _{12})|j\rangle _a,
\end{eqnarray}
where states with subindex $a$ is the ancilla.
Really this is the optimal universal cloning machine presented in Ref.\cite{d_dim}.
For qubit case, we have the well known Bu\v{z}ek-Hillery cloning machine,
\begin{eqnarray}
|0\rangle \rightarrow \sqrt {\frac {2}{3}}|00\rangle |0\rangle _a+\sqrt {\frac {1}{6}}(|01\rangle _{12}+|10\rangle )|1\rangle _a,
\nonumber \\
|1\rangle \rightarrow \sqrt {\frac {2}{3}}|11\rangle |0\rangle _a+\sqrt {\frac {1}{6}}(|01\rangle _{12}+|10\rangle )|0\rangle _a.
\nonumber
\end{eqnarray}

As we already know, besides the case of symmetric output, we can
adjust the qualities of the individual output states in an
imbalanced way. This is realized by the asymmetric cloning machine
\cite{Cerf}. For 1 to 2 unified cloning machine, where projection
$s_2$ is used, we know that, $s_2$ can be written as a summation of
identity and a permutation, $s_2=\frac {1}{2}\left(
\mathbb{I}^{\otimes 2}+\mathcal{P}\right)$, where $\mathcal{P}$ is
the permutation (SWAP) operator, $\mathcal{P}|jl\rangle =|lj\rangle
$. We can then consider to adjust the weights of identity and
permutation in an imbalanced way. Naturally, we can change symmetric
projector to asymmetric case as, $s_2\rightarrow \alpha
\mathbb{I}^{\otimes 2}+\beta \mathcal{P}$, where $\alpha $ and
$\beta $ are weights for adjusting. The corresponding asymmetric
unified cloning machine is now changing as,
\begin{eqnarray}
|\varphi \rangle \rightarrow \alpha |\varphi \rangle _{1}|\Phi
^+\rangle _{2a}+\beta |\varphi \rangle _2|\Phi ^+\rangle _{1a},
\label{asymmetry}
\end{eqnarray}
note the orders of the subindices in these two terms are different,
also those two terms are not orthogonal. The problem now is whether
this cloning procession is optimal. We know that, $|\varphi \rangle
_2|\Phi ^+\rangle _{1a}=\frac {1}{d}\sum (U_{jl}|\varphi \rangle
_1)|\Phi _{jl}\rangle _{2a}$, where $U_{jl}$ are generalized Pauli
matrices and identity, $|\Phi _{jl}\rangle _{2a}$ are orthonormal
maximally entangled states with $|\Phi _{00}\rangle =|\Phi ^+\rangle
$. Now exactly, we find that (\ref{asymmetry}) is the optimal
asymmetric cloning proposed by Cerf \cite{Cerf,Cerfd}.

So far only limited cases of the asymmetric cloning machine have
been presented \cite{Cerf,Cerfd,Cerf3}. The general asymmetric
cloning is still absent possibly because that the formulae are too
complicated to be extended. Here similar as for the case of 1 to 2,
the unified cloning machine can be adjusted to the general
asymmetric cloning machine and the related entanglement sharing
inequalities \cite{singlet}. The method is to plug into a weight for
each essential permutation to modify the symmetric operator $s_M$ in
(\ref{unified}), the problem is like to put $N$ balls into $M$ boxes
with a weight for each choice. Thus we offer a simple realization of
the asymmetric cloning. When all weights are equal, it reduces to
the symmetric case.

\emph{Conclusions}.---We present a unified optimal universal cloning
machine. The cloning procession is equivalent with Werner cloning
machine \cite{Werner} and the one proposed by Fan \emph{et al.}
\cite{Fan} and can be easily adjusted to asymmetric cloning machines
\cite{Cerf,Cerfd,Cerf3,singlet} and to the general case. This simple
cloning machine is always realized by a symmetric projection and
initially prepared maximally entangled states and thus should reduce
the difficulties for implementation. Also the general fidelities are
obtained. Our result offers a new platform for other cloning tasks
for cases like phase-covariant and state-dependent.

This work is supported by NSFC (10974247, 11047174),``973'' program
(2010CB922904) and NFFTBS (J1030310).


\begin{thebibliography}{99}
\bibitem{combination}The equation can be written as,
\begin{eqnarray}
&&\sum _{m=0}^{M-N}\left( \frac {(N+m)!}{N!m!}\right) \left(
\frac {(M-N-m+d-2)!}{(M-N-m)!(d-2)!}
\right)(N+m)
\nonumber \\
&=&N\sum _{m=0}^{M-N}C^N_{N+m}C^{d-2}_{M-N-m+d-2}+
\nonumber \\
&&+(N+1)\sum _{m=0}^{M-N}C_{N+m}^{N+1}C^{d-2}_{M-N-m+d-2}
\nonumber \\
&=&N\cdot C_{M+d-1}^{M-N}+(N+1)C_{M+d-1}^{M-N-1}.
\nonumber
\end{eqnarray}
\end{thebibliography}
\end{document}